\documentclass[floatfix,aps,prb,superscriptaddress,10pt]{revtex4-2}

\usepackage[utf8]{inputenc}
\usepackage{amsmath}
\usepackage{amssymb}
\usepackage{graphicx}
\usepackage{dcolumn}
\usepackage{bm}
\usepackage{chemformula}
\usepackage{setspace}
\usepackage[caption=false]{subfig}
\DeclareUnicodeCharacter{2212}{-}

\def\C60{A$_x$C$_{60}$}

\def\HgCu3{HgCa$_2$Cu$_3$O$_{8+y}$}
\def\HgCu4{HgBa$_2$Ca$_3$Cu$_4$O$_{10+y}$}
\def\TlCu{Tl$_2$Ba$_2$CuO$_{6+\delta}$}
\def\TlCu3{Tl$_2$Ba$_2$Ca$_2$Cu$_3$O$_{10+y}$}
\def\TlCu4{Tl$_2$Ba$_2$Ca$_3$Cu$_4$O$_{12+y}$}

\def\BiCu3{Bi$_2$Sr$_2$Ca$_{2}$Cu$_3$O$_y$}

\def\8LSCO{La$_{1.88}$Sr$_{.12}$CuO$_4$}
\def\110LNSCO{La$_{1.5}$Nd$_{0.4}$Sr$_{0.1}$CuO$_{4}$}
\def\stage4LCO{La$_{2}$CuO$_{4+\delta}$}
\def\Y248{YBa$_2$Cu$_4$O$_8$}

\def\NbSe2{NbSe$_2$}
\def\TaSe2{TaSe$_2$}
\def\TiSe2{TiSe$_2$}
\def\NaCoOH2O{Na$_{0.3}$CoO$_{2y}$H$_2$O}
\def\MgB2{MgB${}_2$}

\def \agl {|\alpha_{GL}|}

\def \tilnem {\tilde{\eta}}
\def \geohat {\hat{\lambda}_1}
\def \bichat {\hat{\lambda}_2}

\begin{document}

\title{Vortex flow anisotropy in nematic  superconductors}

\author{F. Castillo Menegotto}
\affiliation{Universidad de Buenos Aires, Facultad de Ciencias Exactas y Naturales, Departamento de Física. Ciudad Universitaria, 1428, Buenos Aires, Argentina.}
\affiliation{CONICET - Universidad de Buenos Aires, Instituto de Física de Buenos Aires (IFIBA). Buenos Aires, Argentina}
\author{R. S. Severino}
\affiliation{Universidad de Buenos Aires, Facultad de Ciencias Exactas y Naturales, Departamento de Física. Ciudad Universitaria, 1428, Buenos Aires, Argentina.}
\affiliation{CONICET - Universidad de Buenos Aires, Instituto de Física de Buenos Aires (IFIBA). Buenos Aires, Argentina}
\author{P. D. Mininni}
\affiliation{Universidad de Buenos Aires, Facultad de Ciencias Exactas y Naturales, Departamento de Física. Ciudad Universitaria, 1428, Buenos Aires, Argentina.}
\affiliation{CONICET - Universidad de Buenos Aires, Instituto de Física Interdisciplinaria y Aplicada (INFINA). Buenos Aires, Argentina}
\author{E. Fradkin}
\affiliation{Department of Physics and Institute for Condensed Matter Theory, University of Illinois  at Urbana-Champaign,
1110 West Green Street, Urbana, Illinois 61801-3080, USA}
\author{V. Bekeris}
\affiliation{Universidad de Buenos Aires, Facultad de Ciencias Exactas y Naturales, Departamento de Física. Ciudad Universitaria, 1428, Buenos Aires, Argentina.}
\affiliation{CONICET - Universidad de Buenos Aires, Instituto de Física de Buenos Aires (IFIBA). Buenos Aires, Argentina}
\author{G. Pasquini}
\affiliation{Universidad de Buenos Aires, Facultad de Ciencias Exactas y Naturales, Departamento de Física. Ciudad Universitaria, 1428, Buenos Aires, Argentina.}
\affiliation{CONICET - Universidad de Buenos Aires, Instituto de Física de Buenos Aires (IFIBA). Buenos Aires, Argentina}
\author{G. S. Lozano}
\affiliation{Universidad de Buenos Aires, Facultad de Ciencias Exactas y Naturales, Departamento de Física. Ciudad Universitaria, 1428, Buenos Aires, Argentina.}
\affiliation{CONICET - Universidad de Buenos Aires, Instituto de Física de Buenos Aires (IFIBA). Buenos Aires, Argentina}

\begin{abstract}
    We investigate the vortex flow anisotropy in the mixed state of nematic superconductors, focusing on the effects of nematic-superconducting coupling on vortex dynamics. Using numerical simulations within a time-dependent Ginzburg-Landau (TDGL) approach, we analyze vortex viscosity in a model featuring an s-wave superconducting order parameter coupled to an Ising-like nematic order parameter, suitable for systems with $C_4$ symmetry. Our results indicate that nematicity induces a significant anisotropy in the flux-flow resistivity, which depends on both vortex core shape anisotropy and normal-phase conductivity anisotropy. These two effects can either compete or cooperate with each other. We discuss the implications of these findings for identifying nematic superconductivity in the superconducting phase. Our work provides new insights into the interplay between nematic and superconducting order parameters, leading to new possibilities for experimental and theoretical exploration of anisotropic transport properties in unconventional superconductors.
\end{abstract}

\maketitle

\section{Introduction}

Electronic nematicity \cite{kivelson1,Fradkin_2010}, first documented in experiments in two-dimensional electron gases (2DEG) in magnetic fields \cite{kivelson2,lilly-1999}, is observed in both cuprate \cite{Ando_2002, Hinkov_2008, Comin_2015} and iron-based superconductors \cite{Chuang_2010, Prozorov_2009, Kuo_2016, Chu_2010, Kuo_2012, Gallais_2013, Tanatar_2016, Kretzschmar_2016}, and has been theoretically proposed as a crucial factor underlying unconventional superconductivity in these materials \cite{Fradkin_2010, Fradkin_2015, Fernandes_2022}. Although a consensus on the details of the microscopic origin of the pairing mechanism responsible for superconductivity in these compounds is yet to be reached, there are strong  hints of the significant role played by an electronic nematic order that interacts with the superconducting order in either a cooperative or competitive manner, depending on the considered material \cite{Prozorov_2009, Kuo_2016, Fente2018, Li_2017, Lederer2017, Kalisky_2010, Kalisky_2011, Fente2018, zhang_2019, Fernandes_2022, Klem_2024}. In this context, there is strong numerical evidence that strong quantum electronic nematic fluctuations can lead to superconductivity in quantum Monte Carlo simulations of a relatively simple model system \cite{Lederer2017}. At a phenomenological level, this interplay can be explored under a Ginzburg-Landau formalism in terms of a coupling between the nematic and superconducting order parameters  \cite{Fradkin_2010, chowdhury}.

Experimentally, nematic order manifests in different ways. The signature is an anisotropic normal phase with a breaking of the tetragonal $C_4$ symmetry in both structural and transport properties \cite{Chuang_2010, Chu_2010, Tanatar_2010}. It is now well established that this symmetry breaking is driven by electronic degrees of freedom, consistent with the existence of a genuine nematic phase \cite{Chu2012, Kuo_2012, Kuo_2016}. A large resistivity anisotropy develops in the nematic phase up to the superconducting transition, although its quantitative determination is in some cases a practical challenge due, among other factors, to the formation of nematic domains  \cite{sanches_nature, Bartlett_2021}. 

Below the superconducting critical temperature, $T_c$, the interplay between superconductivity and nematicity is not  thoroughly understood. One of the unresolved questions is whether a nematic superconducting state arises as a result of this interaction. In this framework, the mixed superconducting state offers an interesting playground to explore  these questions. In type II superconductors, vortices penetrate the samples, so both transport and magnetic properties are mainly determined by vortex dynamics. Nematic superconductivity would also be reflected in an anisotropic vortex shape \cite{Severino2022}. In fact, superconducting vortices with an elliptical core have been observed in FeSe compounds \cite{song, Lu_2018, zhang_2019}, consistent with the presence of a nematic superconducting order parameter. In contrast, no detectable in-plane anisotropy in the superconducting stiffness has been reported in optimally doped BaFeCoAs samples \cite{schmidt}.

A key observable to investigate the nematic-superconducting interplay is the resistivity anisotropy in the mixed superconducting phase. The electrical resistivity of superconductors arises from the motion of vortices as a result of the presence of an average current density. In the simplest standard model of vortex dynamics, developed by Bardeen and Stephen (BS) \cite{Bardeen1965}, a vortex is treated as a normal core with a circular cross section of radius equal to the coherence length, which experiences a Lorentz force per unit length, $\vec{F}_L = (\phi_0/c)\vec{J} \times \hat{e}_z$. Here, $\phi_0$ is the flux quantum, $\vec{J}$ is the current density, $\hat{e}_z$ is a unit vector in the direction of the magnetic field and $c$ is the speed of light. In the pure flux flow regime this Lorentz force is balanced by a viscous dissipative force, $\vec{F}_d = -\nu_{BS} \vec{v}$. In its simplest formulation, the BS viscosity coefficient is given by $\nu_{BS} = \phi_0^2/(2 \pi \xi^2 c^2 \rho_n)$, where $\xi$ denotes the superconducting coherence length and $\rho_n$ is the normal-state resistivity. The flux-flow resistivity is inversely proportional to this viscosity coefficient \cite{stephen1965, kim1965}.

Since its original formulation, the derivation of the BS viscosity coefficient has been revisited by many authors. The BS force is fundamental when modeling vortex dynamics as a dynamical system of interacting particles subjected to viscous drag, in the presence of disorder and thermal fluctuations. This approach has proven useful in explaining the various dynamical regimes of  vortex matter, such as DC and AC depinning transitions \cite{daroca2010}, as well as the existence of glassy phases. It has been applied to both conventional and unconventional superconductors \cite{Blatter1994}. The Bardeen-Stephen estimate was extended to account for anisotropic superconductors by Hao and Clem in 1991 \cite{hao1991}, by incorporating anisotropy both in the normal conductivity tensor and in the superconducting stiffness. However, this analytical approach did not consider any coupling between these anisotropies.

The purpose of this work is to analyze from a theoretical perspective the effects of the nematic-superconducting coupling on the transport properties in the mixed superconducting phase.  For this end, we employ a Time-Dependent Ginzburg-Landau (TDGL) model for a nematic superconductor, with an $s$-wave superconducting order parameter coupled to an Ising-like nematic order parameter. Our main aim is to evaluate the applicability of this simple anisotropic model to quantify the interplay between superconducting and nematic orders, and the resistivity anisotropy.

This work is organized as follows. In Section~\ref{sec:TDGL}, we revisit the simplest TDGL model for an anisotropic superconductor, derive its fundamental equations and key assumptions. In Section~\ref{sec:viscosity}, we describe how this model can be adapted to calculate the viscosity, providing a detailed explanation of our methodology and presenting the main results of our numerical simulations. Finally, in Section~\ref{sec:conclusions}, we summarize our findings and discuss their implications for the study of nematicity in the superconducting phase, highlighting the potential significance of our results for its characterization.

\section{Flux flow viscosity in the Time Dependent Ginzburg Landau formalism}\label{sec:TDGL}
The aim of this work is to study the effects of flux-flow anisotropy in nematic superconductors within the TDGL formulation of a nematic superconductor. The formulation is discussed in detail in \cite{Severino2022,Severino2024}. The part of the free energy for a complex $s$-wave superconductor order parameter $\psi$ (related to the superfluid Cooper pair density via  $|\psi|^2 = n_s$) is 
\begin{equation}
    F_S = \int_V \left[ \alpha_{GL}|\psi|^2 + \frac{\beta_{GL}}{2}|\psi|^4 + \frac{\hbar^2}{2m}|\mathcal{D}\psi|^2 + \frac{(\nabla\times\mathbf{A})^2}{8\pi} \right],
    \label{eq:free_energy}
\end{equation}
where $\boldsymbol{A}$ is the electromagnetic vector potential related to the magnetic induction via $\nabla\times\boldsymbol{A} = \boldsymbol{B}$, and  $\alpha_{GL}$ and $\beta_{GL}$ are the standard temperature-dependent Ginzburg-Landau parameters. In particular $\alpha_{GL} = \alpha_0(T-T_c)$, which changes sign at $T = T_c$, signaling the phase transition from the normal to the superconductor state. The differential operator $\mathcal{D} = -i\nabla - \frac{e}{\hbar c}$ is the standard covariant derivative, where $e$ denotes the charge of the Cooper pairs, $\hbar$ is the reduced Planck constant and $c$ is the speed of light. We note that while $m$ is a parameter with units of mass, it does not represent the mass of any relevant particle but can be related to the stiffness of the order parameter. In this work we will consider configurations with translational invariance in the $\hat{e}_z$ direction, so any dependence on this coordinate will be hereby ignored. 

In contrast to classical fluids, where the nematic order parameter has tensor-like properties, the coupling to the underlying tetragonal lattice reduces the nematic order parameter to be represented by a real Ising-type scalar field $\eta$ with free energy 
\begin{equation}
    F_N = \int_V \Big[\gamma_2 (\nabla \eta)^2 + \gamma_3 \eta^2 + \frac{\gamma_4}{2} \eta^4 \Big].
\end{equation}
The coupling between nematicity and superconductivity will be included in the model via two terms: a biquadratic coupling between the SC and the nematic order parameters and a trilinear coupling which explicitly breaks the symmetry of the system. For the biquadratic term, we write 
\begin{equation}
    F_{bi} = \lambda_2 \int_V \eta^2 |\psi|^2 dV.
\end{equation}
As remarked before, this term is not an intrinsic nematic term, as it is not related \textit{per se} to the breaking of any spatial symmetry. A term of this type could be present in several theories with multiple order parameters.  The sign of $\lambda_2$ does play an important role in setting whether the nematic and superconducting orders compete or cooperate.

We also introduce a trilinear term of the form
\begin{equation}  
   F_{tr} = \frac{\hbar}{2m}\lambda_1 \int_V \eta e_{ij}\mathcal{D}_i\psi  (\mathcal{D}_j \psi)^* ,
\end{equation}
where $e_{ij}=2 (n_i n_j-\frac{1}{2}\delta_{ij})$ and $\mathbf{n}=(\cos\alpha,\sin\alpha)$ is a vector signaling the orientation of nematicity with respect to the chosen coordinate system. In several types of unconventional superconductors that exhibit nematic behavior (such as Fe-based and some cuprates) a structural phase transition from tetragonal to orthorhombic is observed, with a relative angle of $\pi/4$ between vector basis of the lattice. The $ab$ plane of the crystal structure (the $\hat{e}_x$ and $\hat{e}_y$ directions in our system) can be chosen to align with the tetragonal or the orthorhombic axes, which will then select a value for $\alpha$. In this work, we will align the $ab$ plane with the \textit{orthorhombic} axes, which implies $\alpha = 0$. We will comment on the value of $\lambda_1$ and its sign later.

For the superconductor and nematic order parameters, we will consider standard dissipative dynamics (model A in the terminology of Ref. \cite{Hohenberg-1977})
\begin{equation}
    \frac{\hbar^2}{2mD}\partial_t \psi = -\frac{\delta F}{\delta \psi^{*}} , \qquad
    \frac{\hbar^2}{2m D_n} \partial_t \eta = - \frac{\delta F}{\delta \eta} .
\end{equation}
In these expressions, $D$ and $D_n$ are diffusion constants. A derivation for $D$ has been given for BCS superconductors in the original paper by Schmid on TDGL \cite{schmid}. However, we are not aware of realistic estimates of this quantity for non-conventional superconductors, and the same applies for $D_n$. 

 We rescale the order parameters as, $ \psi = \sqrt{\rho_0}\tilde{\psi}$ and $ \eta = \eta_0 \tilde{\eta}$, where $\eta_0^2=-\gamma_3/\gamma_4$ and $\rho_0={|\alpha_{GL}|}/{\beta_{GL}}$.  We also define dimensionless parameters, $\geohat={{\lambda}_1\eta_0}/{\hbar}$, $\bichat={\lambda_2\eta_0^2}/{\agl}$ and $\Gamma_4={\gamma_4\eta_0^4}/({\agl \rho_0})$ (for more details, see Ref. \cite{Severino2022}). There are three relevant length scales in our problem, the superconducting coherence length, $\xi^2 = {\hbar^2}/({2m|\alpha_{GL}|})$, the London penetration length, $\lambda_L^2 ={mc^2}/({4\pi e^2 \rho_0})$ and the nematic coherence length, $l_\eta^2={\gamma_2}/{|\gamma_3}|$.

For the electromagnetic vector potential evolution, we will consider a modification to the equation used in \cite{Severino2022, Severino2024},
\begin{equation}
    \frac{\boldsymbol{\rho}^{-1}}{c^2} \partial_t \boldsymbol{A} = -\frac{\delta F}{\delta \boldsymbol{A}},
    \label{eq:ohm}
\end{equation}
where $\boldsymbol{\rho}$ is now the resistivity {\em tensor}
\begin{equation}
\rho_{ij}=\rho_n(\delta_{ij} + (\delta-c_1 \tilde{\eta} )e_{ij}).
\end{equation}
If we take the coordinate system as the {\em orthorhombic} one, then
\begin{equation}
    \boldsymbol{\rho}=\rho_n ( \boldsymbol{I}+
    (\delta- c_1 \tilde{\eta}) \boldsymbol{\sigma_3}) ,
    \label{eq:rho}
\end{equation}
where $\boldsymbol{\sigma}_3$ is the Pauli matrix. Here  $\delta$ is a dimensionless parameter related to an intrinsic anisotropy, and $c_1$ characterizes the nematic resistivity anisotropy. Notice that we are assuming that the resistivity anisotropy of the normal phase is linear in the nematic order parameter (for a discussion see \cite{sanchez2021}). In what follows, we will assume that  $\delta = 0$. In principle, the resistivity tensor also has off-diagonal terms that would be relevant if we were interested in studying Hall effects. These effects will not be considered here.
The relevant time scales for the dynamics of each field are given by $\tau_{r_\psi}=\frac{\xi^2}{2D}$,
$\tau_{r_n}=\frac{l_\eta^2}{2D_n}$ and $\tau_{r_A}=\frac{2\pi\lambda_L^2}{\rho_nc^2}$, with the ratio of characteristic times for $\psi$ and $A$  given by $\tau_{r_A}=(l_E^2/\xi^2) \tau_{r_\psi}$, where $l_E$ is the electric field penetration depth.

Within this framework, the superconducting current density, as derived from the dynamical equation for the vector potential (see Ref.~\cite{Severino2022}), is given by
\begin{equation}
\boldsymbol{J}_s = \left( \boldsymbol{I} + \hat\lambda_1 \tilnem \boldsymbol{\sigma}_3 \right) \left( \frac{\hbar e}{2im} \left( \psi^* \nabla \psi - \psi \nabla \psi^* \right) - \frac{e^2}{mc} |\psi|^2 \boldsymbol{A} \right).
\end{equation}
Note that this is not the conventional definition, as it includes an additional term arising from the coupling to the nematic order parameter. On the other hand, in the \textit{zero electrostatic potential gauge}, the dissipative component of the current density reads
\begin{equation}
    \boldsymbol{J}_n = -\frac{\boldsymbol{\rho}^{-1}}{c} \partial_t \boldsymbol{A}.
\end{equation}

The most relevant parameters for the effect we want to study are $\hat{\lambda}_1$ and $c_1$. In our approach, $\eta(x,y,t)$ is a dynamical field coupled to the superconducting order parameter $\psi$ and the electromagnetic field $\boldsymbol{A}$, so that its behavior is dictated by the full solutions of the dynamical equations. However, some behavior can be anticipated on general grounds. Excluding the presence of domain walls, $\eta$ is a smooth field that will take a value close to its uniform expectation value $\tilde{\eta}_v$. Inside a vortex, $\tilde\eta$ can increase (decrease) depending on whether the sign of $\lambda_2$ is positive (negative). 

As shown in Ref. \cite{Severino2022}, the main effect of the trilinear coupling is to produce an anisotropic coherence length that, when considering a constant nematic background, could be thought of as a mass ratio (more properly, an order parameter rigidity ratio),
\begin{equation}
    \frac{\xi^2_b}{\xi^2_a}=\frac{1-\hat{\lambda}_1 \tilde{\eta}_v}{1+\hat{\lambda}_1 \tilde{\eta}_v}=\frac{m_a}{m_b},
    \label{eq:masses}
\end{equation}
where $\xi_{a(b)}$ are the coherence lengths in the $ab$ plane. Thus, $\hat{\lambda}_1$ controls the ellipticity of the vortex core.
On the other hand, by neglecting the small (possible) contribution of the intrinsic anisotropy, from Eq. (\ref{eq:rho})
\begin{equation}
    \frac{\sigma_a}{\sigma_b}=\frac{1+c_1 \tilde{\eta}_v}{1-c_1 \tilde{\eta}_v}.
\label{eq:cinductivities}
\end{equation}
where $\sigma_{a(b)}$ are the normal-state conductivities in the $ab$ plane ($\boldsymbol{\sigma} = \boldsymbol{\rho}^{-1}$). Then, as a first approximation, i.e., ignoring the fact that $\eta$ is indeed a dynamical field, we can think of a nematic superconductor as an anisotropic superconductor with different in-plane ``mass parameters'' and anisotropic normal conductivity given by Eqs. (\ref{eq:masses}), (\ref{eq:cinductivities}). Hao and Clem generalized the BS model for a homogeneous anisotropic superconductor and solved the problem analytically in the limit of infinite $\kappa$ \cite{hao1991}, obtaining an anisotropic viscosity
\begin{eqnarray}
\nu_a&=&\frac{\phi_0 \tilde H_{c2}}{4 c^2}\left(\frac{m_a}{m_b}  \sigma_a + 3 \sigma_b\right),\\
\nu_b&=&\frac{\phi_0 \tilde H_{c2}}{4 c^2}\left(\frac{m_b}{m_a}  \sigma_b + 3 \sigma_a\right),
\end{eqnarray}
where $\tilde H_{c2}=\phi_0/(2\pi\xi_a\xi_b)$ is the upper critical magnetic field. Then, the viscosity anisotropy can be estimated as
\begin{equation}
    \Delta \nu=\frac{\nu_a-\nu_b}{\nu_a+\nu_b}.
    \label{eq:anisviscos}
\end{equation}
The fact that viscosity anisotropy is the combined result of two effects, namely, resistivity anisotropy of the normal phase plus vortex ellipticity, is certainly correct. However, we know that the BS approach yields a very crude approximation even for the case of standard Abrikosov vortices. Moreover, in the case of nematic superconductors, $\eta$ is a dynamical field coupled with the superconducting order parameter. We  will then analyze this phenomenon with a more careful approach by solving the TDGL model, and the resulting degree of flux flow nematicity will be estimated as in Eq. (\ref{eq:anisviscos}).

To express the results in dimensionless form, we adopt the following normalization scheme. Lengths are measured in units of the London penetration depth \( \lambda_L \), while time is scaled by the relaxation time of the order parameter \( \tau_{r_\psi} \). Velocities are thus normalized by a characteristic velocity \( v_{\psi} = \lambda_L / \tau_{r_\psi} \). Magnetic fields are expressed in units of the isotropic upper critical field \( H_{c2} = \phi_0 / (2\pi \xi^2) \), and current densities are given in units of the superconducting critical current \( J_c = c H_{c2} / (4\pi \lambda_L) \).

\section{Numerical study of flux flow viscosity}\label{sec:viscosity}
\subsection{Boundary conditions: The trap and the transport current}\label{sec:trap}

We solve the TDGL equations numerically using pseudo-spectral methods. Key characteristics of these methods are their exponential convergence and the absence of numerical dissipation and dispersion, which make them an accurate tool for the estimation of transport properties. However, this approach comes with a cost due to their reliance on periodic boundary conditions within the simulation domain $(x,y)$ of dimensions $2\pi L\times 2 \pi L$ (the system is assumed to be translationally invariant along the $z$-axis). At first glance, this requirement appears to be in conflict with the need for non-zero circulation of the vector potential $\boldsymbol{A}$ in vortex configurations, as vortices typically involve phase singularities and quantized flux. However, this issue can be resolved, e.g.,  by employing a technique involving images of the system within the simulation box. For a detailed explanation of these ideas and their application to this specific problem, see Ref. \cite{Severino2022}.

\begin{figure}[ht]
      \includegraphics[width=0.5\columnwidth]{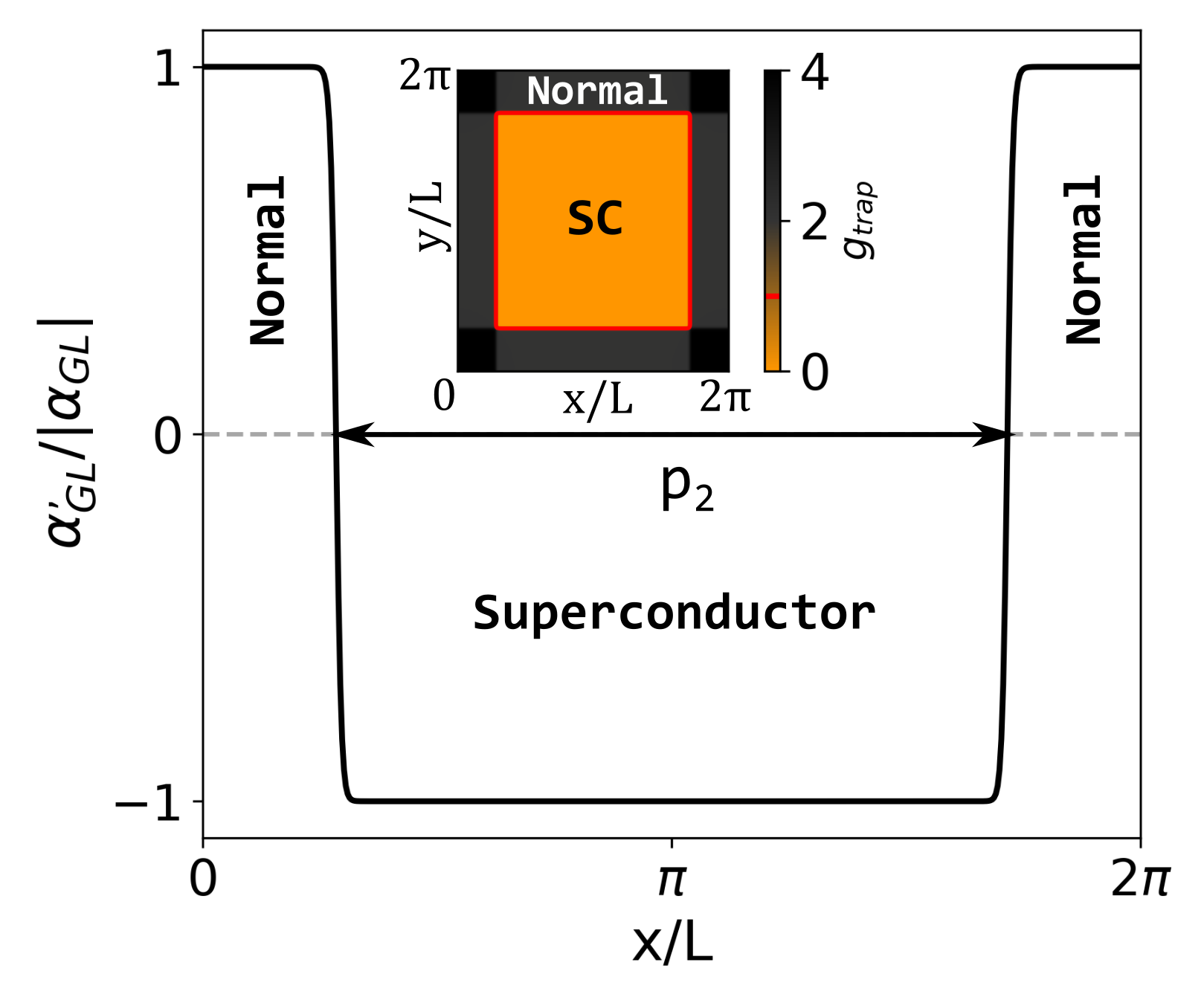}
 \caption{Cross section of the effective Ginzburg–Landau parameter \( \alpha_{GL}' / |\alpha_{GL}| \) for the trap function \( g_{\text{trap}}(x, y = \pi) \) in Eq. (\ref{eq: trap func}), with parameters \( p_1 = 0.07 \) (inverse of the transition slope), \( p_2 = 4.5 \) (superconductor size), and \( V_0 = 2 \) (amplitude). The inset shows a density plot of the trapping potential function over the simulation domain. The superconducting region (\( \alpha_{GL}' < 0 \)) is shown in orange, while the normal region (\( \alpha_{GL}' > 0 \)) appears in gray and black. The red line indicates the boundary between the two regions, defined by the condition \( \alpha_{GL}' = 0 \).}
   \label{fig: trap}
\end{figure}

In the present work, we will instead use a different approach inspired by an idea already used in numerical studies of vortices in superfluids \cite{super_fluid_bec}. To this end, we introduce a {\em potential trap} that confines the superconducting order parameter to a certain region $\mathcal{R}$ within our simulation box. This can be done by adding to the free energy an additional term of the form
\begin{equation}
    F_{trap} = - \int_V  \alpha_{GL} g_{trap}(x,y) |\psi|^2 dV,
\end{equation}
where $g_{trap}$ is a dimensionless function that describes the spatial dependence of the trapping potential. To be compatible with the numerical method, the function $g_{trap}$ must satisfy periodic boundary conditions. For our purposes, it is sufficient to ask that
\begin{equation}
g_{trap}(x,y) < 1 \quad \text{if} \quad (x,y) \in \mathcal{R}, \qquad \qquad  g_{trap}(x,y) > 1 \quad \text{if} \quad (x,y) \notin \mathcal{R}.
\end{equation}
In this way, neglecting proximity effects, inside  (outside) $\mathcal{R}$ the material is in the superconducting (normal) state. In this work, we use 
\begin{equation}\label{eq: trap func}
    g_{trap}(x,y) = V_0(f(x) + f(y)),
\end{equation}
where $V_0$ is a parameter that controls the amplitude of the trapping potential, and $f$ is the step-like function defined by
\begin{equation}
\label{eq: f function}
    f(x)=\frac{1}{2} \left\lbrace 1 + \tanh\left(\frac{\cos[\frac{x}{L}] + \cos[\frac{p_2}{2}]}{\frac{p_1}{2}\sin[\frac{p_2}{2}]}\right) \right\rbrace,
\end{equation}
with parameters $p_1$ and $p_2$ that controls the edge sharpness and size of the trap, respectively. A plot of the resulting trap for the parameter values used in this work, $p_1 = 0.07$, $p_2 = 4.5$ and $V_0=2$ can be seen in Fig. \ref{fig: trap}. A simple way to understand the effect of this term is as follows. Consider the quadratic term in the superconducting order parameter,
\begin{equation}
    \alpha_{GL}' |\psi|^2= \alpha_{GL} (1- g_{trap}(x,y))|\psi|^2.
\end{equation}
 If $g_{trap}(x,y) \simeq 0 $ then $\alpha_{GL}' \simeq \alpha_{GL}$ ($<0$), which implies that the region is in the superconducting state. Instead, if $g_{trap}(x,y) > 1 $, then $\alpha_{GL}'>0$ so the sample is in the normal state. Therefore, this implementation allows us to easily create a normal-superconductor interface. For the specific problem we are considering, the details of the  trap potential will not impact on the final results.

In addition, we need to introduce an external current $\boldsymbol{J}_{ext}$, which is the origin of the driving force related to the vortex motion. The free energy is then modified to include the work provided by the current source, by means of the Legendre transformation $G=F+\frac{1}{c} \boldsymbol{J}_{ext}\cdot\boldsymbol{A}$. Therefore, the dynamic equation to solve is now
\begin{equation}
\label{energia con J}
    \frac{\boldsymbol{\rho}^{-1}}{c} \partial_t \boldsymbol{A} = -c\frac{\delta F}{\delta \boldsymbol{A}}-\boldsymbol{J}_{ext}.
\end{equation}
With this condition, in the stationary regime, the total current flowing across the sample, including both superconducting and normal density currents, is provided by the external source.

As previously mentioned, due to the nature of pseudo-spectral methods, the variables must be periodic functions within the simulation domain. Consequently, the current density must not only be periodic itself but also the curl of a periodic function,  in order to guarantee the periodicity of the magnetic field. This implies that the total current across the simulation domain must integrate to zero. In more physical terms, this ensures that all current flowing in must also flow out. For this purpose, in the present work, we use
\begin{equation}
\label{Jext}
    \boldsymbol{J}_{ext}(x,y) = J_0(f(x)-\zeta)\hat{e}_y \quad \text{or} \quad \boldsymbol{J}_{ext}(x,y) = J_0(f(y)-\zeta)\hat{e}_x ,
\end{equation}
depending on the desired direction of current flow, where $f$ is the function defined in Eq. (\ref{eq: f function}), but evaluated with different values of $p_1$ and $p_2$. We fix the constant $\zeta$ such that $\int_0^{2\pi L} d\theta (f(\theta)-\zeta)=0$. We have taken $p_1=0.05$ and $p_2 = 5.8$. The parameter \( J_0 \) sets the amplitude of the external driving current. With this choice, we obtain a current that is predominantly uniform in the central region and exhibits a strong backflow at the edges, outside the superconducting region (see Fig. \ref{fig: scheme}). 

The main differences between these equations and those used in Refs. \cite{Severino2022,Severino2024}, are the inclusion of the trapping potential, the conditions imposed under external current biasing, and the terms proportional to $c_1$ that account for the conductivity anisotropy in the normal phase.

\subsection{Simulations and results}\label{results}

We now proceed with the numerical evaluation of vortex viscosity in the mixed phase. The use of pseudo-spectral methods for solving TDGL equations in nematic superconductors has been described in \cite{Severino2022,Severino2024}. 

We are interested in the behavior of $\nu_{lg}$ and $\nu_{sh}$ as a function of the vortex ellipticity, quantified by its aspect ratio,
\begin{equation}\label{eq: aspect ratio}
a_R=\frac{\xi_{long}}{\xi_{short}}=\sqrt{\frac{1+|\hat{\lambda}_1 \tilde{\eta}_v|}{  1-|\hat{\lambda}_1 \tilde{\eta}_v|    }},
\end{equation} 
and the normal-state resistivity anisotropy $c_1$. To carry out the numerical simulations, it is necessary to specify the model parameters. For the characteristic length scales, we set $\xi=0.06 L$, $l_\eta=1.5\xi$ ($\Gamma_4=1$), and $\lambda_L=2\xi$ ($\kappa=2$). Regarding the time scales, we take $\tau_{r_\eta}\simeq2.12 \tau_{r_\psi}$ (so that $D_n \simeq 1.06 D $), and choose $\rho_n$ such that $\tau_{r_A}\simeq 0.68 \tau_{r_\psi}$. The system is studied in the weak biquadratic coupling limit, i.e., $\hat{\lambda}_2 = 0$, in which case $\tilnem_v=1$. The simulation domain, of size $2\pi L \times 2\pi L$ (with $L = 1$), is discretized into an $N \times N$ numerical grid with $N = 256$, and time evolution is performed with a time step $\Delta t \simeq 4.71 \times 10^{-3} \tau_{r_\psi}$. With these parameters, the size of the trap (as given in Sec. \ref{sec:trap}) is 37.5 $\lambda_L$.

\begin{figure}[ht]
    \subfloat[\label{fig: scheme}]{
      \includegraphics[width=0.38\columnwidth]{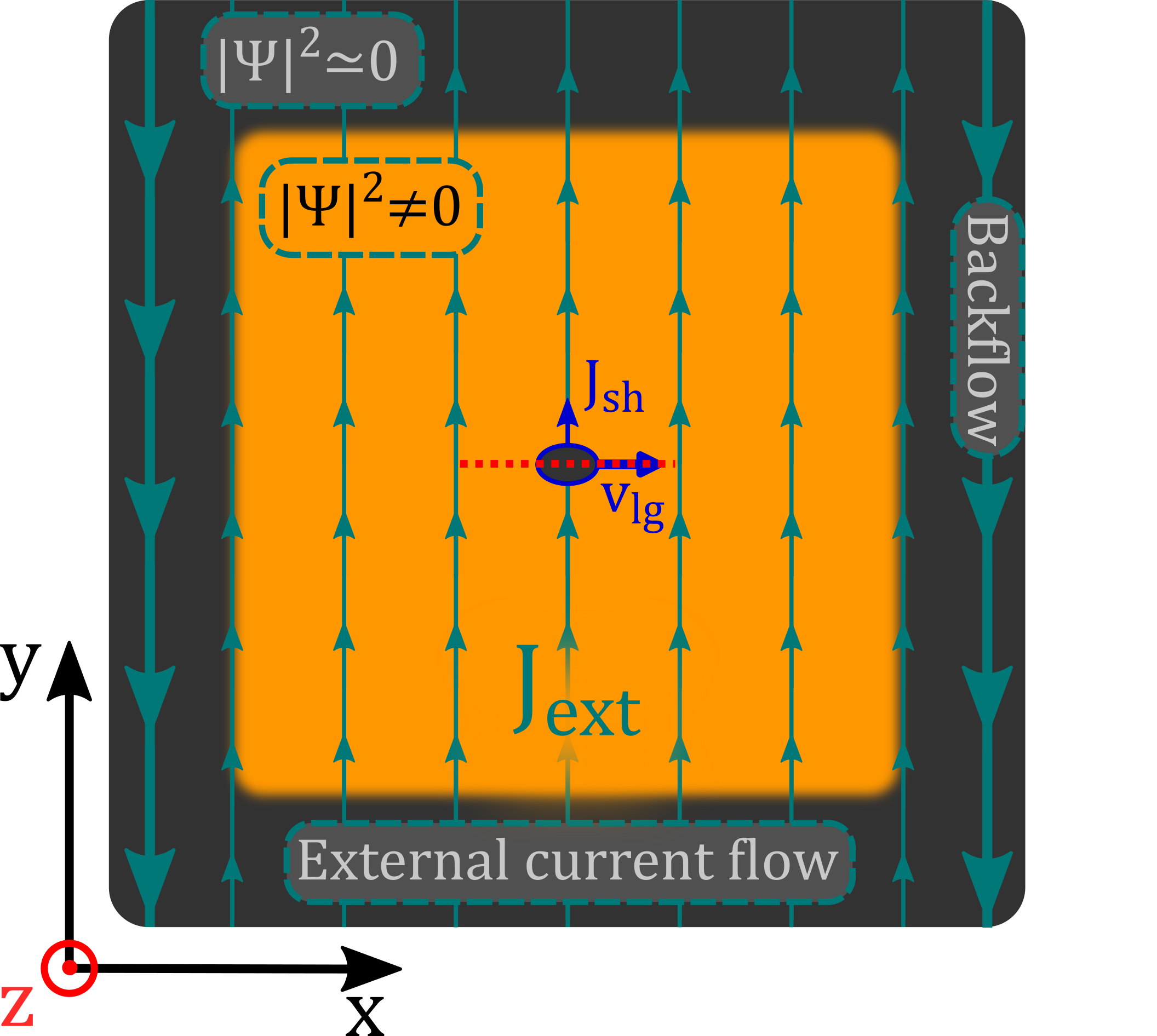}
    }
    \hfill
    \subfloat[\label{fig: current profiles}]{
      \includegraphics[width=0.49\columnwidth]{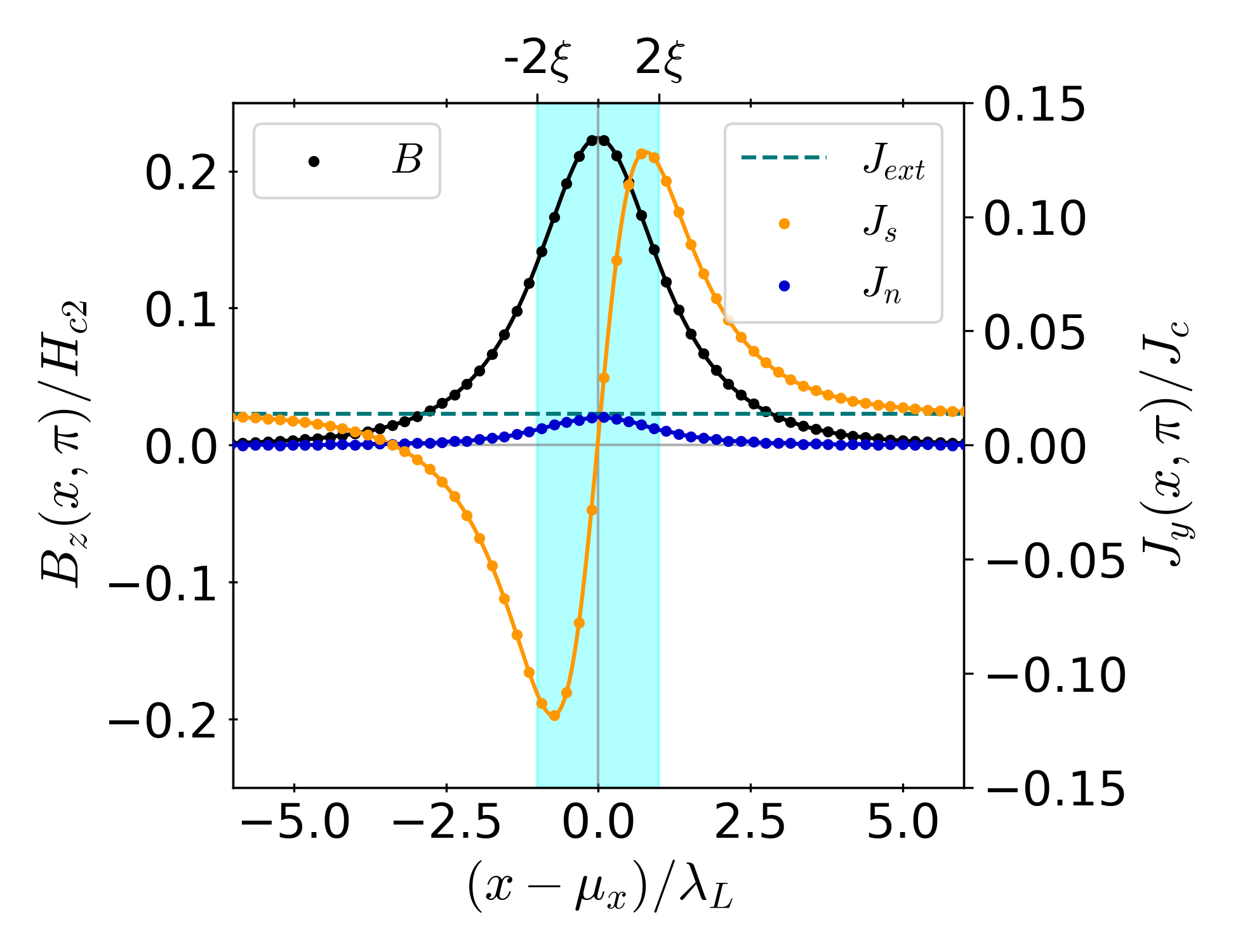}
    }
    \caption{(a) Schematic illustration of vortex flow. The superconducting (normal) region is shown in orange (gray). The external current $\boldsymbol{J}_{\text{ext}}$ (green arrows) is approximately uniform inside the superconducting region and exhibits a backflow in the normal region to ensure zero net flow across the system. In this case, $\boldsymbol{J}_{\text{ext}}$ is aligned with the short axis of the ellipse and is thus denoted as $J_{sh}$, so the vortex moves in the direction of its long axis with velocity $v_{lg}$ (the magnetic induction $B$ points in the $\hat{e}_z$ direction). (b) Cross-sectional profiles of the normal ($J_n$, blue dots), superconducting ($J_s$, orange dots), and external ($J_{\text{ext}}$, green dashed line) current density components (right axis), together with the magnetic induction $B_z$ (left axis, black dots), taken along the section $y/L=\pi$ indicated by red dots in panel (a). The profiles are centered around the vortex center position $\mu_x(t)$ at a given time $t$ in the stationary regime. Solid lines are guides to the eye. This representative simulation was performed with $a_R\simeq 1.53$.}
    \label{fig:esquema}
\end{figure}

To begin, we choose initial conditions in such a way that, after a short evolution, a single elliptical vortex (with one magnetic flux $\phi_0$) is formed at the center of the simulation box. We then turn on $\boldsymbol{J}_{\text{ext}}$, which is approximately uniform within the superconductor. Inside the normal core of the vortex, this external drive gives rise to a dissipative current component, which constitutes the actual mechanism responsible for the vortex motion. Its direction is chosen in such a way that the current density inside the sample coincides either with the {\em long} axis ($J_{lg}$) or the {\em short} one ($J_{sh}$) of the ellipse. As a result, the vortex moves perpendicularly to that direction. Fig. \ref{fig: scheme} provides a schematic representation of the case where the external current is applied along the short axis of the ellipse, causing the vortex to move along its long axis. This configuration allows us to extract the viscosity along the long direction, $\nu_{lg}$. In this setup, the magnetic field associated with the vortex points along the $z$-axis, directed outward from the plane of the screen.

In Fig. \ref{fig: current profiles}, we show the cross-section profiles of the normal ($J_n$), superconducting ($J_s$), and external ($J_{\text{ext}}$) current density components along the direction transverse to the vortex motion ($y$-axis), together with the magnetic induction $B_z$, evaluated around the vortex core. In the core region (where $|\tilde\psi|^2\approx 0$), the normal current density dominates, satisfying $\boldsymbol J_n \simeq \boldsymbol J_{\text{ext}}$ both in magnitude and direction. Far from the vortex, the normal current density decays to zero, and the superconducting current density becomes predominant, with $\boldsymbol J_s \simeq \boldsymbol J_{\text{ext}}$.

\begin{figure}[ht]
    \subfloat[\label{fig: vortex_track}]{
      \includegraphics[width=0.45\columnwidth]{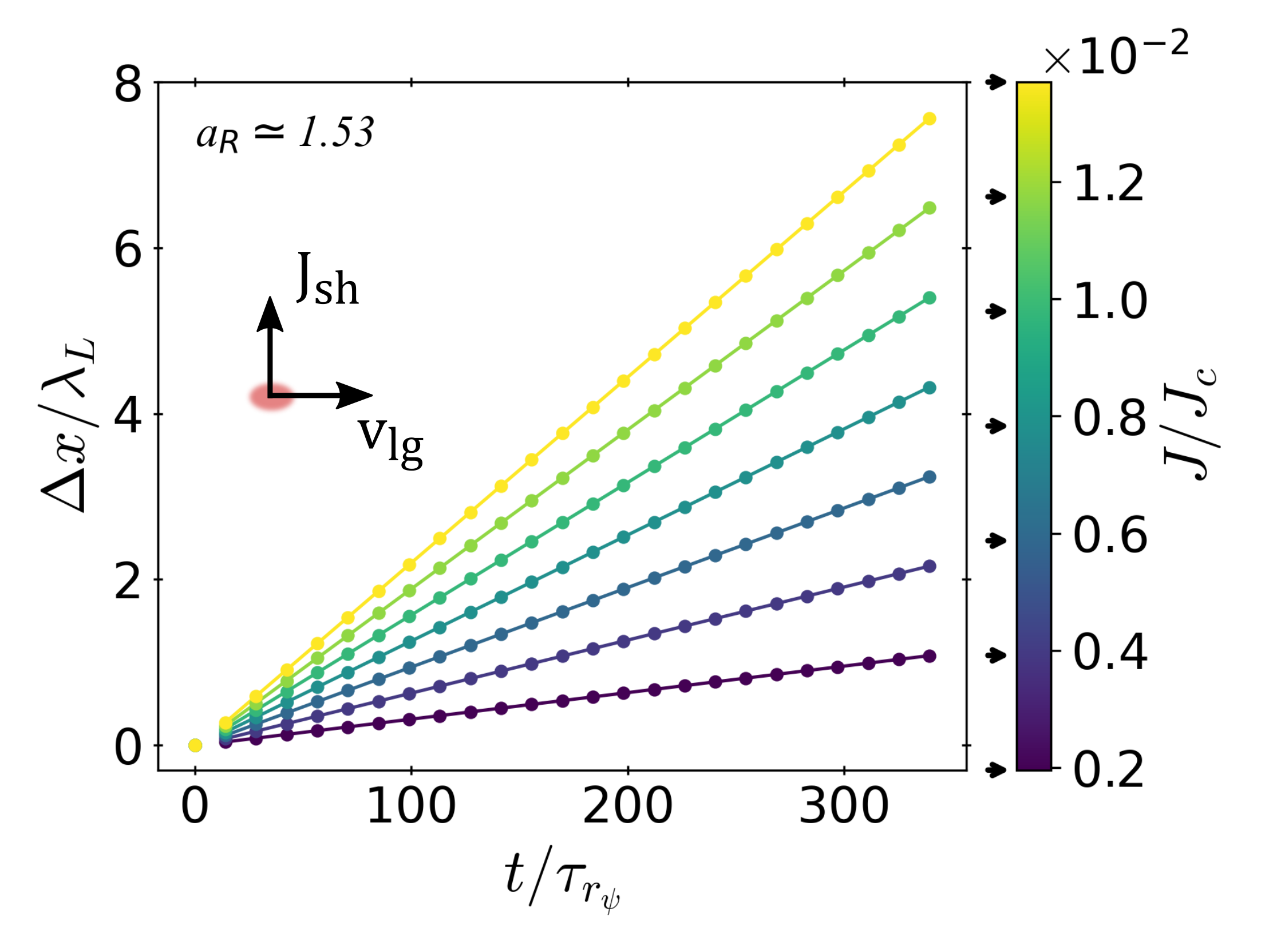}
    }
    \hfill
    \subfloat[\label{fig: vortex_velocities}]{
      \includegraphics[width=0.45\columnwidth]{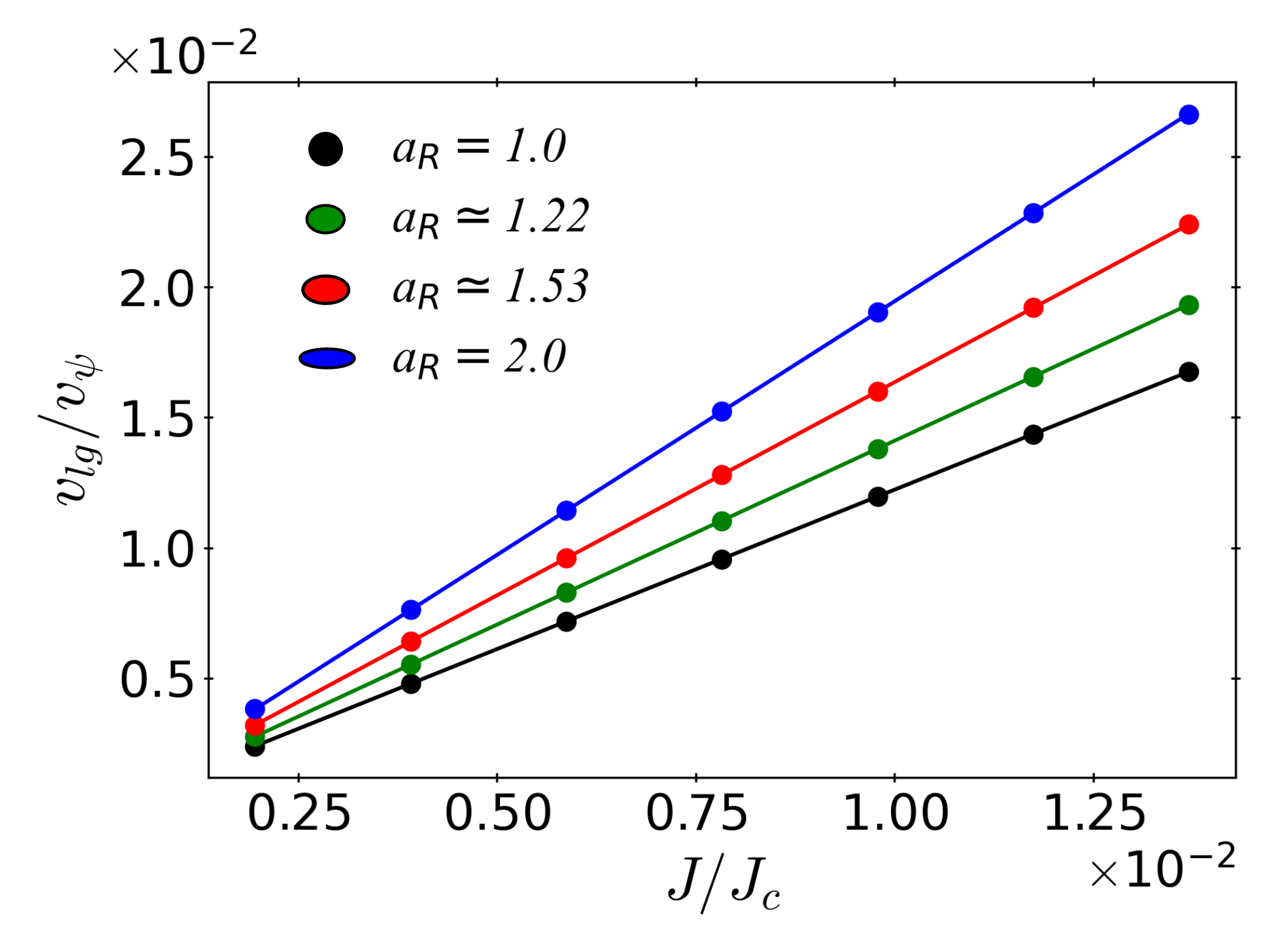}
    }
    \caption{(a) Vortex displacement of an elliptical vortex with $a_R\simeq1.53$ and $c_1=0$ as a function of time, for different external current amplitudes (all other parameters fixed, see text). Dots represent the tracked vortex position, and the solid line shows the linear fit of the trajectory. Arrows indicate applied current, ${J}_{sh}$, along the short axis of the elliptical vortex, and the vortex velocity, $v_{lg}$, along the long axis. (b) Drift velocity $v_{lg}$ as a function of the external driving current for different ellipticities, $\geohat = 0.0$, $0.2$, $0.4$ and $0.6$ (black, green, red and blue dots). The slope obtained from each linear fit (solid lines) is proportional to the inverse of the viscosity (Eq.~\ref{nul}).}
    \label{fig: visosity calculation}
\end{figure}

\begin{figure}[ht]
    \subfloat[\label{fig: results A}]{
      \includegraphics[width=0.45\columnwidth]{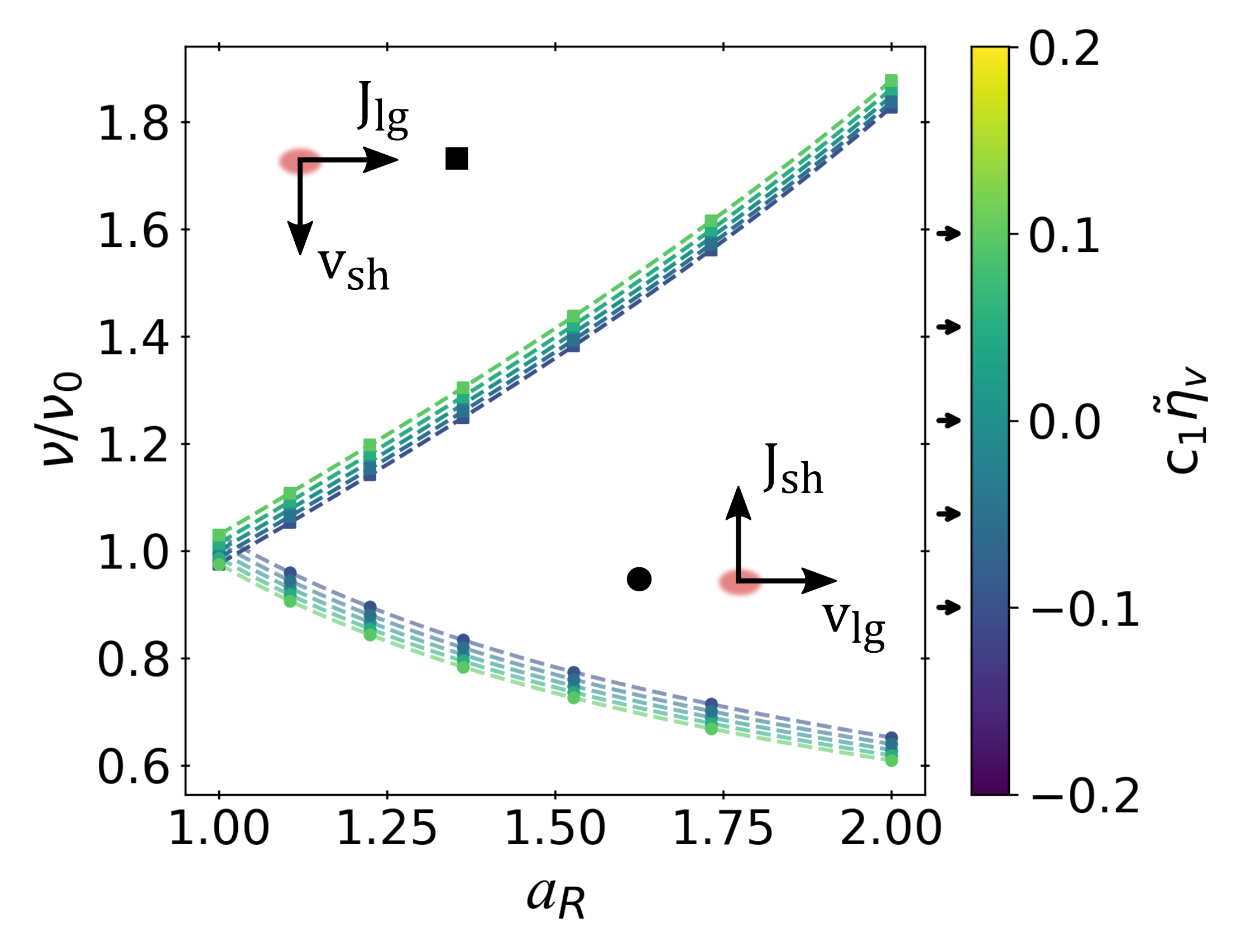}
    }
    \hfill
    \subfloat[\label{fig: results B}]{
      \includegraphics[width=0.45\columnwidth]{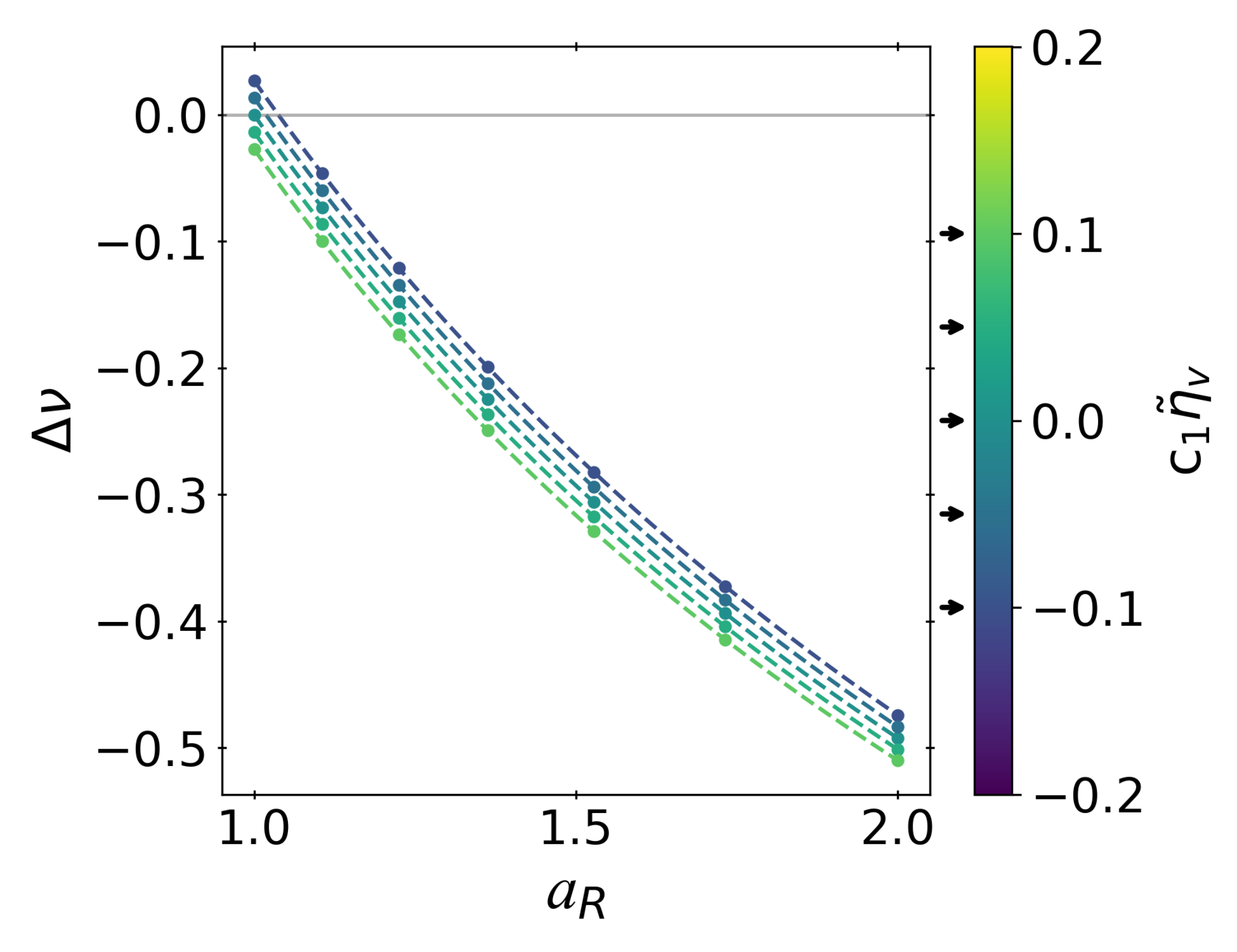}
    }
    \caption{(a) Viscosity $\nu_{sh}$ ({\tiny\(\blacksquare\)}) and $\nu_{lg}$ (\(\bullet\)) as a function of the aspect ratio $a_R$, for five different values of $c_1$, with a color scheme marked on the right. These results are normalized by the viscosity coefficient $\nu_0/\nu_{BS}\simeq 0.60$, which is obtained in the isotropic case, i.e., with $\geohat = c_1=0$. (b) The viscosity anisotropy $\Delta \nu=(\nu_{lg}-\nu_{sh})/(\nu_{lg}+\nu_{sh})$ as a function of the aspect ratio $a_R$ and for five different values of $c_1=\lbrace 0,\pm 0.05,\pm 0.1 \rbrace$. Notice that for negative values of $c_1$, $\Delta \nu$ changes sign, indicating that the geometric factor and the resistivity anisotropy of the normal state compete in these cases. Dashed lines are guides to the eye.}
    \label{fig: results}
\end{figure}

To determine the numerical value of $\nu_{lg}$, we track the position of the vortex as a function of time. Examples of this procedure are shown in Fig. \ref{fig: vortex_track} for specific values of the parameters of interest, $\geohat=0.4$ (which, using Eq. (\ref{eq: aspect ratio}), leads to $a_R\simeq 1.53$) and $c_1=0$, and for different values of the external current density. From the slope of the trajectories, we are then able to determine the drag velocity, $v_{lg}$, corresponding to each current. In Fig. \ref{fig: vortex_velocities}, the velocity $v_{lg}$ is plotted as a function of $J_{sh}$ for different values of $a_R$. Using the flux flow relationship 
\begin{equation}
   v_{lg}=\left(\frac{\phi_0 }{\nu_{lg}c}\right)J_{sh}, 
    \label{nul}
\end{equation}
we can extract the value of $\nu_{lg}$ using a simple linear fit. This procedure was performed for different values of {$a_R$}, considering both positive and negative values of $c_1$, and applying current in both directions, $J_{sh}$ and $J_{lg}$. The results are summarized in Fig. \ref{fig: results A}, while  Fig. \ref{fig: results B} shows the corresponding viscosity anisotropies $\Delta \nu$. Notably, when $c_1 < 0$, both geometrical and normal-state resistivity effects reinforce each other, whereas for $c_1 > 0$ they compete, leading to a specific case where these effects cancel out exactly. Moreover, at least for the data shown, the magnitude of $\Delta \nu$ is comparable to that of $a_R$, suggesting that this effect could be experimentally measurable.

We conclude this section with some remarks and discuss concrete examples of real materials. One of the most well-studied Fe-based superconductors is the FeSe compound. In this material, clear vortex images were obtained, for instance, in Ref. \cite{putilov2019} using STM technique. It can be seen in Fig. 1 of that reference that vortices are aligned with the orthorhombic axes, in one of the Fe-Fe bond directions. From the same figure, an estimate of the aspect ratio yields approximately $a_R \approx 1.7$.

However, we note that the literature does not show a clear correlation between the sign of $c_1$ (which indicates which lattice axis is more resistive) and the orthorhombic distortion. For instance, Ref. \cite{Bartlett_2021} reports that the longest orthorhombic axis is more resistive than the shortest one from $90$ K down to $40$ K, but the trend then reverses from $40$ K down to $T_c \sim 9$ K. On the other hand, Ref. \cite{Tanatar_2016} does not report this sign change.  Therefore, the correlation between the sign of resistivity anisotropy and the vortex core alignment is still under debate.

In the case of the BaFeCoAs family we are not aware of any vortex imaging studies reporting ellipticity, which makes it unclear whether techniques employed lack sufficient resolution to observe it, or if the trilinear coupling is absent or very small in these materials, as suggested by transport experiments in strained optimally doped compounds \cite{schmidt}. 

Additionally, a reliable estimation of the intrinsic resistivity anisotropy in the normal orthorhombic phase of these materials is still lacking. The interpretation of experimental data is complicated, partly due to the multiple sources of resistivity contributions (such as domain walls), and also due to the huge elasto resistivity and the difficulty in controlling the absolute sample strain. Nevertheless, as a reasonable estimate, resistivity anisotropies of up to $10$\% (in absolute value) seem plausible in FeSe \cite{Tanatar_2016}.

\section{Conclusions}
\label{sec:conclusions}
 
{\em Free} flux flow is one of the most fundamental mechanisms of transport in the superconducting phase, although it corresponds to a highly idealized physical situation. In a realistic case, transport is influenced by various types of disorder, thermal fluctuations, and vortex-vortex interactions. While it is not known whether all these effects are simply additive, a widely used theoretical approach to studying transport involves molecular dynamics simulations, where vortices are treated as point particles governed by the equation  
\begin{equation}
    \boldsymbol{F}_{i,vv} + \boldsymbol{F}_{i,dis} + \boldsymbol{F}_{i,T} + \boldsymbol{F}_{ext} + \boldsymbol{F}_{i,\nu} = 0,
\end{equation}
where the different terms represent the force exerted on the $i$-vortex due to, respectively, vortex-vortex interactions, interactions with disorder, thermal fluctuations, the force exerted by an external current, and finally the viscous drag, given by $\boldsymbol{F}_{i,\nu} = -\nu \boldsymbol{v}_i$.  Although this approach is highly simplified, it provides valuable qualitative insights into the complex dynamical regimes characterizing vortex matter and transport in realistic superconductors.  

In this paper we investigated vortex motion in nematic superconductors using the TDGL formalism and numerically analyzed the viscosity tensor. 
Our results indicate that a minimal yet essential modification of this approach, in the presence of nematic order, requires replacing the scalar drag coefficient $\nu$ with a tensor $\boldsymbol{\nu}$ in the drag force.

As we discussed in previous sections, the tensor character of $\boldsymbol\nu$ originates from two sources: a geometrical factor (the ellipticity of the vortex core), generated by the superconducting-nematic coupling, and the conductivity anisotropy of the normal phase. It would also be interesting to explore the existence of non-diagonal terms related to Hall physics. An even more realistic model would incorporate an additional {\em dumbbell} variable to capture possible fluctuations of nematicity along the sample and account for other sources of anisotropy of the vortex core. Models of this type have recently attracted considerable attention in the context of active matter, where it is well known that dynamical phases are generically richer than their symmetric counterparts. We hope to report on these issues in the near future.
 
Our results demonstrate that nematicity can induce pronounced transport anisotropies in the flux-flow regime, providing a clear and definitive signature of nematic order in the mixed phase of Fe-based superconductors. This establishes a powerful criterion for identifying nematic superconductivity through transport measurements in the superconducting phase. 

We point out that nematic domain walls also dramatically modify vortex physics in the superconducting state \cite{Blatter1994}. In the last decade, new devices have been developed \cite{Hicks2014, Barber2018} and employed \cite{sanches_nature, schmidt, Bartlett_2021} to carefully detwin samples by applying controlled uniaxial stress and strain at low temperatures, obtaining a single nematic domain. This opens a way to explore intrinsic vortex flux flow viscosity anisotropy in the superconducting mixed state.

Furthermore, while the interplay between elastic deformation and anisotropy has been extensively studied in the normal phase, few studies have addressed this topic in the superconducting phase. Our findings suggest that a similar interplay in the superconducting phase could manifest in different observable properties. For example, $a_R$ could be actively controlled via elastic deformation. If this is the case, such tuning could provide a pathway for the experimental detection of elasto-flux-flow resistivity anisotropy.  

In summary, these results open new opportunities for tuning and probing nematic superconductors, offering a direct means to manipulate and characterize their anisotropic transport properties.

\section*{Acknowledgments}

This work was supported in part by the University of Buenos Aires (UBACyT 20020220100129BA), Foncyt (PICT Raices 2019-015890) and CONICET (PIP 112202101004760), Argentina, as well as by the US National Science Foundation through grant No. DMR 2225920 at the University of Illinois, USA. We thank Mariano Marziali Bermúdez for useful comments, and EF thanks Rafael Fernandes for discussions. 

\bibliographystyle{unsrt}
\bibliography{nem}

\begin{thebibliography}{10}

\bibitem{kivelson1}
S.~A. Kivelson, E.~Fradkin, and V.~J. Emery.
\newblock {Electronic liquid-crystal phases of a doped Mott insulator}.
\newblock {\em Nature}, 393:550, 1998.

\bibitem{Fradkin_2010}
E.~Fradkin, S.~A. Kivelson, M.~J. Lawler, J.~P. Eisenstein, and A.~P. Mackenzie.
\newblock {Nematic Fermi Fluids in Condensed Matter Physics}.
\newblock {\em Annual Review of Condensed Matter Physics}, 1(1):153--178, 2010.

\bibitem{kivelson2}
E.~Fradkin and S.~A. Kivelson.
\newblock {Liquid-crystal phases of quantum Hall systems}.
\newblock {\em Phys. Rev. B}, 59:8065, 1999.

\bibitem{lilly-1999}
M.~P. Lilly, K.~B. Cooper, J.~P. Eisenstein, L.~N. Pfeiffer, and K.~W. West.
\newblock {Evidence for an Anisotropic State of Two-Dimensional Electrons in High Landau Levels}.
\newblock {\em Phys. Rev. Lett.}, 82:394--97, 1999.

\bibitem{Ando_2002}
Yoichi Ando, Kouji Segawa, Seiki Komiya, and A.~N. Lavrov.
\newblock Electrical resistivity anisotropy from self-organized one dimensionality in high-temperature superconductors.
\newblock {\em Phys. Rev. Lett.}, 88:137005, 2002.

\bibitem{Hinkov_2008}
V.~Hinkov, D.~Haug, B.~Fauqué, P.~Bourges, Y.~Sidis, A.~Ivanov, C.~Bernhard, C.~T. Lin, and B.~Keimer.
\newblock Electronic liquid crystal state in the high-temperature superconductor {YBa$_2$Cu$_3$O$_{6.45}$}.
\newblock {\em Science}, 319(5863):597--600, 2008.

\bibitem{Comin_2015}
R.~Comin, R.~Sutarto, E.~H. da~Silva~Neto, L.~Chauviere, R.~Liang, W.~N. Hardy, D.~A. Bonn, F.~He, G.~A. Sawatzky, and A.~Damascelli.
\newblock Broken translational and rotational symmetry via charge stripe order in underdoped \ch{Y Ba_{2} Cu_{3} O_{6+y}}.
\newblock {\em Science}, 347(6228):1335--1339, 2015.

\bibitem{Chuang_2010}
T.-M. Chuang, M.~P. Allan, Jinho Lee, Yang Xie, Ni~Ni, S.~L. Bud’ko, G.~S. Boebinger, P.~C. Canfield, and J.~C. Davis.
\newblock Nematic electronic structure in the "parent" state of the iron-based superconductor \ch{Ca(Fe_{1-x}Co_{x})_{2} As_2}.
\newblock {\em Science}, 327(5962):181--184, 2010.

\bibitem{Prozorov_2009}
R.~Prozorov, M.~A. Tanatar, N.~Ni, A.~Kreyssig, S.~Nandi, S.~L. Bud'ko, A.~I. Goldman, and P.~C. Canfield.
\newblock Intrinsic pinning on structural domains in underdoped single crystals of \ch{Ba(Fe_{1-x} Co_{x})_2 As_{2}}.
\newblock {\em Phys. Rev. B}, 80:174517, 2009.

\bibitem{Kuo_2016}
H.~Kuo, J.~Chu, J.~C. Palmstrom, S.~A. Kivelson, and I.~R. Fisher.
\newblock Ubiquitous signatures of nematic quantum criticality in optimally doped $\ch{Fe}$-based superconductors.
\newblock {\em Science}, 352:958, 2016.

\bibitem{Chu_2010}
Jiun-Haw Chu, James~G. Analytis, Kristiaan~De Greve, Peter~L. McMahon, Zahirul Islam, Yoshihisa Yamamoto, and Ian~R. Fisher.
\newblock In-plane resistivity anisotropy in an underdoped iron arsenide superconductor.
\newblock {\em Science}, 329(5993):824--826, 2010.

\bibitem{Kuo_2012}
H.~H. Kuo, J.~G. Analytis, J.~H. Chu, R.~M. Fernandes, J.~Schmalian, and I.~R. Fisher.
\newblock Magnetoelastically coupled structural, magnetic, and superconducting order parameters in \ch{BaFe_2(As_{1-x} P_x)_2}.
\newblock {\em Phys. Rev. B}, 86:134507, 2012.

\bibitem{Gallais_2013}
Y.~Gallais, R.~M. Fernandes, I.~Paul, L.~Chauvi\`ere, Y.-X. Yang, M.-A. M\'easson, M.~Cazayous, A.~Sacuto, D.~Colson, and A.~Forget.
\newblock Observation of incipient charge nematicity in \ch{Ba(Fe_{1-x}Co_{x})_2 As_{2}}.
\newblock {\em Phys. Rev. Lett.}, 111:267001, 2013.

\bibitem{Tanatar_2016}
M.~A. Tanatar, A.~E. B\"ohmer, E.~I. Timmons, M.~Sch\"utt, G.~Drachuck, V.~Taufour, K.~Kothapalli, A.~Kreyssig, S.~L. Bud'ko, P.~C. Canfield, R.~M. Fernandes, and R.~Prozorov.
\newblock Origin of the resistivity anisotropy in the nematic phase of \ch{FeSe}.
\newblock {\em Phys. Rev. Lett.}, 117:127001, 2016.

\bibitem{Kretzschmar_2016}
F.~Kretzschmar, T.~Böhm, U.~Karahasanovic, B.~Muschler, A.~Baum, D.~Jost, J.´ Schmalian, S.~Caprara, M.~Grilli, C.~Di Castro, J.~G. Analytis, J.-H. Chu, I.~R. Fisher, and R.~Hackl.
\newblock Critical spin fluctuations and the origin of nematic order in \ch{Ba (Fe_{1−x} Co_x)_2 As_{2}}.
\newblock {\em Nature Physics}, 12:560, 2016.

\bibitem{Fradkin_2015}
Eduardo Fradkin, Steven~A. Kivelson, and John~M. Tranquada.
\newblock Colloquium: Theory of intertwined orders in high temperature superconductors.
\newblock {\em Rev. Mod. Phys.}, 87:457--482, 2015.

\bibitem{Fernandes_2022}
R.~F. Fernandes, A.~I. Coldea, H.~Ding, I.~R. Fisher, P.~J. Hirschfeld, and G.~Kotliar.
\newblock Iron pnictides and chalcogenides: a new paradigm for superconductivity.
\newblock {\em Nature}, 601:35, 2022.

\bibitem{Fente2018}
A.~Fente, A.~Correa-Orellana, A.~E. B\"ohmer, A.~Kreyssig, S.~Ran, S.~L. Bud'ko, P.~C. Canfield, F.~J. Mompean, M.~Garc\'{\i}a-Hern\'andez, C.~Munuera, I.~Guillam\'on, and H.~Suderow.
\newblock {Direct visualization of phase separation between superconducting and nematic domains in Co-doped ${\mathrm{CaFe}}_{2}{\mathrm{As}}_{2}$ close to a first-order phase transition}.
\newblock {\em Phys. Rev. B}, 97:014505, Jan 2018.

\bibitem{Li_2017}
J.~Li, P.~J. Pereira, J.~Yuan, Y-Y. Lv, M-P. Jiang, D.~Lu, Z-Q. Lin, Y-J. Liu, J-F. Wang, L.~Li, X.~Ke, Van~T. G., M-Y. Li, H-L. Feng, T.~Hatano, H-B. Wang, P-H. Wu, K.~Yamaura, E.~Takayama-Muromachi, J.~Vanacken, L.~F. Chibotaru, and V.~V. Moshchalkov.
\newblock Nematic superconducting state in iron pnictide superconductors.
\newblock {\em Nat. Commun..}, 8:1880, 2017.

\bibitem{Lederer2017}
Lederer S., Schattner Y., Berg E., and Kivelson~S. A.
\newblock {Superconductivity and non-Fermi liquid behavior near a nematic quantum critical point}.
\newblock {\em Proc. Natl. Acad. Sci.}, 114(19):4905--4910, 2017.

\bibitem{Kalisky_2010}
B.~Kalisky, J.~R. Kirtley, J.~G. Analytis, Jiun-Haw Chu, A.~Vailionis, I.~R. Fisher, and K.~A. Moler.
\newblock Stripes of increased diamagnetic susceptibility in underdoped superconducting $\ch{Ba(Fe_{1-x} Co_{x})_{2} As_{2}}$ single crystals: Evidence for an enhanced superfluid density at twin boundaries.
\newblock {\em Phys. Rev. B}, 81:184513, 2010.

\bibitem{Kalisky_2011}
B.~Kalisky, J.~R. Kirtley, J.~G. Analytis, J.-H. Chu, I.~R. Fisher, and K.~A. Moler.
\newblock Behavior of vortices near twin boundaries in underdoped $\ch{Ba (Fe_{1-x} Co_{x})_2 As_{2}}$.
\newblock {\em Phys. Rev. B}, 83:064511, 2011.

\bibitem{zhang_2019}
Irene~P. Zhang, Johanna~C. Palmstrom, Hilary Noad, Logan Bishop-Van~Horn, Yusuke Iguchi, Zheng Cui, Eli Mueller, John~R. Kirtley, Ian~R. Fisher, and Kathryn~A. Moler.
\newblock Imaging anisotropic vortex dynamics in \ch{FeSe}.
\newblock {\em Phys.Rev. B}, 100:024514, 2019.

\bibitem{Klem_2024}
M.~L. Klemm, SH. Mozaffari, R.~Zhang, B.~W. Casas, A.~E. Koshelev, M.~Yi, L.~Balicas, and P.~Dai.
\newblock Nematic superconductivity from selective orbital pairing in iron pnictide single crystals.
\newblock {\em Cell Rep. Phys. Sci.}, 5:101816, 2024.

\bibitem{chowdhury}
D.~Chowdhury, E.~Berg, and S.~Sachdev.
\newblock Nematic order in the vicinity of a vortex in superconducting \ch{FeSe}.
\newblock {\em Phys. Rev.B}, 84:205113, 2011.

\bibitem{Tanatar_2010}
M.~A. Tanatar, E.~C. Blomberg, A.~Kreyssig, M.~G. Kim, N.~Ni, A.~Thaler, S.~L. Bud'ko, P.~C. Canfield, A.~I. Goldman, I.~I. Mazin, and R.~Prozorov.
\newblock Uniaxial-strain mechanical detwinning of ${\text{cafe}}_{2}{\text{as}}_{2}$ and ${\text{bafe}}_{2}{\text{as}}_{2}$ crystals: Optical and transport study.
\newblock {\em Phys. Rev. B}, 81:184508, May 2010.

\bibitem{Chu2012}
Jiun-Haw Chu, Hsueh-Hui Kuo, James~G. Analytis, and Ian~R. Fisher.
\newblock Divergent nematic susceptibility in an iron arsenide superconductor.
\newblock {\em Science}, 337(6095):710--712, 2012.

\bibitem{sanches_nature}
J.~J. Sanchez, P.~Malinowski, J.~Mutch, J.~Liu, J.~W. Kim, P.~J. Ryan, and J.~Chu.
\newblock The transport–structural correspondence across the nematic phase transition probed by elasto x-ray diffraction.
\newblock {\em Nat. Mater.}, 20:1519, 2021.

\bibitem{Bartlett_2021}
J.~M. Bartlett, A.~Steppke, S.~Hosoi, H.~Noad, J.~Park, C.~Timm, T.~Shibauchi, A.~P. Mackenzie, and C.~W. Hicks.
\newblock Relationship between transport anisotropy and nematicity in \ch{FeSe}.
\newblock {\em Phys. Rev. X}, 11:021038, 2021.

\bibitem{Severino2022}
R.~S. Severino, P.~D. Mininni, V.~Bekeris E.~Fradkin, G.~Pasquini, and G.~S. Lozano.
\newblock Vortices in a ginzburg-landau theory of superconductors with nematic order.
\newblock {\em Phys. Rev. B}, 106:094512, 2022.

\bibitem{song}
C.~L. Song, Y.~L. Wang, P.~Cheng, Y.~P. Jiang, W.~Li, T.~Zhang, Z.~Li, K.~He, L.~Wang, and J.~F.~Jia et~al.
\newblock Direct observation of nodes and twofold symmetry in fese superconductor.
\newblock {\em Science}, 332:1410, 2011.

\bibitem{Lu_2018}
D.~Lu, Y.~Lv, J.~Li, B.~Zhu, Q.~Wang, H.~Wang, and P.~Wu.
\newblock Elliptical vortex and oblique vortex lattice in the fese superconductor based on the nematicity and mixed superconducting order.
\newblock {\em npj Quantum Materials}, 3(12), 2018.

\bibitem{schmidt}
J.~Schmidt, V.~Bekeris, G.~S. Lozano, M.~V. Bortule, M.~Marziali Bermudez, C.~W. Hicks, P.~C. Canfield, E~Fradkin, and G.~Pasquini.
\newblock Nematicity in the superconducting mixed state of strain detwinned underdoped $\ch{Ba (Fe_{1-x} Co_{x})_2 As_{2}}$.
\newblock {\em Phys. Rev.B}, 99:064515, 2019.

\bibitem{Bardeen1965}
John Bardeen and M.~J. Stephen.
\newblock Theory of the motion of vortices in superconductors.
\newblock {\em Phys. Rev.}, 140:A1197--A1207, Nov 1965.

\bibitem{stephen1965}
M.~J. Stephen and J.~Bardeen.
\newblock Viscosity of type-ii superconductors.
\newblock {\em Phys. Rev. Lett.}, 14:112--113, Jan 1965.

\bibitem{kim1965}
Y.~B. Kim, C.~F. Hempstead, and A.~R. Strnad.
\newblock Flux-flow resistance in type-ii superconductors.
\newblock {\em Phys. Rev.}, 139:A1163--A1172, Aug 1965.

\bibitem{daroca2010}
D.~Pérez Daroca, G.~S. Lozano, G.~Pasquini, and V.~Bekeris.
\newblock Depinning and dynamics of ac driven vortex lattices in random media.
\newblock {\em Phys. Rev. B}, 81:184520, 2010.

\bibitem{Blatter1994}
G.~Blatter, M.~V. Feigel'man, V.~B. Geshkenbein, A.~I Larkin, and V.~M. Vinokur.
\newblock Vortices in high-temperature superconductors.
\newblock {\em Rev. Mod. Phys.}, 66:1125, 1994.

\bibitem{hao1991}
Z.~Hao and J.~R. Clem.
\newblock Viscous flux motion in anisotropic type-2 superconductors in low fields.
\newblock {\em IEEE Trans.Magn.}, 27:1086, 1991.

\bibitem{Severino2024}
R.~S. Severino, P.~D. Mininni, V.~Bekeris E.~Fradkin, G.~Pasquini, and G.~S. Lozano.
\newblock Ginzburg-landau approach to the vortex–domain wall interaction in superconductors with nematic order.
\newblock {\em Phys. Rev. B}, 109:094513, 2024.

\bibitem{Hohenberg-1977}
P.~C. Hohenberg and B.~I. Halperin.
\newblock Scaling laws for dynamic critical phenomena.
\newblock {\em Reviews of Modern Physics}, 49:435, 1977.

\bibitem{schmid}
A.~Schmid.
\newblock A time dependent {Ginzburg-Landau} equation and its application to the problem of resistivity in the mixed state.
\newblock {\em Phys. Kondens Materie}, 5:302, 1966.

\bibitem{sanchez2021}
J.~J. Sanchez and P.~Malinowskiand J.~Mutch et~al.
\newblock The transport–structural correspondence across the nematic phase transition probed by elasto x-ray diffraction.
\newblock {\em Nat Mater}, 20:1519–1524, 2021.

\bibitem{super_fluid_bec}
Julian Amette~Estrada, Marc~E. Brachet, and Pablo~D. Mininni.
\newblock Turbulence in rotating bose-einstein condensates.
\newblock {\em Phys. Rev. A}, 105:063321, Jun 2022.

\bibitem{putilov2019}
A.~V. Putilov, C.~Di~Giorgio, V.~L. Vadimov, D.~J. Trainer, E.~M. Lechner, J.~L. Curtis, M.~Abdel-Hafiez, O.~S. Volkova, A.~N. Vasiliev, D.~A. Chareev, G.~Karapetrov, A.~E. Koshelev, A.~Yu. Aladyshkin, A.~S. Mel'nikov, and M.~Iavarone.
\newblock Vortex-core properties and vortex-lattice transformation in fese.
\newblock {\em Phys. Rev. B}, 99:144514, Apr 2019.

\bibitem{Hicks2014}
Clifford~W. Hicks, Mark~E. Barber, Stephen~D. Edkins, Daniel~O. Brodsky, and Andrew~P. Mackenzie.
\newblock Piezoelectric-based apparatus for strain tuning.
\newblock {\em Review of Scientific Instruments}, 85(6):065003, 06 2014.

\bibitem{Barber2018}
Mark~E. Barber, Alexander Steppke, Andrew~P. Mackenzie, and Clifford~W. Hicks.
\newblock Piezoelectric-based uniaxial pressure cell with integrated force and displacement sensors.
\newblock {\em Review of Scientific Instruments}, 90(2):023904, 02 2019.

\end{thebibliography}
\end{document}